# Detector of alpha particles and x-rays operating in ambient air in pulse counting mode or/and with gas amplification

G. Charpak[1], P. Benaben[1], P. Breuil[1], V.Peskov[1,2]
[1]Ecole Nationale Superieure des Mines de St. Etienne, France
[2]CERN, Geneva, Switzerland

**Abstract**

Ionization chambers working in ambient air in current detection mode are attractive due to their simplicity and low cost and are widely used in several applications such as smoke detection, dosimetry, therapeutic beam monitoring and cetera. The aim of this work was to investigate if gaseous detectors can operate in ambient air in pulse counting mode as well as with gas amplification which potentially offers the highest possible sensitivity in some measurements for example for alpha particle detection or high energy x-ray photon or electron detection.
To investigate the feasibility of this method two types of open- end gaseous detectors were build and successfully tested. The first one was a single wire or multiwire cylindrical geometry detector operating in pulse mode at a gas gain of one (pulse ionization chamber). This detector was readout by a custom made wide -band charge sensitive amplifier able to deal with slow induced signals generated by slow motion of negative and positive ions. The multiwire detector was able to detect alpha particles with efficiency close to 22%. The second type alpha detector was an innovative GEM-like detector with resistive electrodes operating in air in avalanche mode at high gas gains (up to $10^4$). This detector can also operate in a cascaded mode or be combined with other detectors, for example with MICROMEGAS. This detector was readout by a conventional charge -sensitive amplifier and was able to detect alpha particles with 100% efficiency. This detector could also detect x-rays photons or fast electrons. A detailed comparison between these two detectors is given as well as comparison with the commercially available alpha detectors. The main advantages of gaseous detectors operating in air in a pulse detection mode are their simplicity, low cost and high sensitivity. One of the possible applications of these new detectors is alpha particle background monitors which, due to their low cost can find wide application not only in houses, but in public areas: airports, railway station and so on.

## I. Introduction

In some applications, like dosimetry (see for example [1]) or therapeutic beam monitoring (see for example [2]) it is very attractive to use gaseous detectors operating in air. Indeed, various designs of ionization chamber operating in air were already developed, built and nowadays are successfully used not only in various research laboratories, but as very useful practical devices. One of the widely used detectors is a smoke detector which one can seen in many houses. [3]. It is a small ionization chamber containing an alpha source; the current produced by the alpha source is continuously measured by the electronic circuit. In the case of appearance the smoke the current drops and this triggers the alarm. Advantages of this approach are simplicity and the cost.
The aim of this work is to investigate if for some applications, for example for alpha particle detection or X-ray detection one can exploit the gaseous detector operating in ambient air in a pulse counting mode - by detecting each individual charge pulse produced by an alpha particle or by an X-ray photon. In contrast to traditional ionization chambers, measuring the ionization current, this approach offers higher sensitivity as well as a capability of evaluating the pulse height spectrum of the radiation.
Of course, operation in pulse mode in air has some difficulties because one has to deal with slow signals induced by slowly moving negative and positive ions and this impose special requirements on the detector geometry and the front-end electronics.
To investigate the feasibility of this method, two types of open- end gaseous detectors were built and successfully tested. The first one was an open-end single wire or multiwire cylindrical geometry detector operating in pulse mode at a gas gain of one (pulse ionization chamber).
The second type alpha detector was an innovative GEM-like detector with resistive electrodes operating in air in avalanche mode at high gas gains (up to $10^4$). This detector can also operate in a cascaded mode or be combined with other detectors, for example with MICROMEGAS.
As will be shown below the second type has much better signal to noise ratio and as a result - higher sensitivity. This allowed us to detect not only individual alpha particles, but also single high energy x-ray photons.
We will report below our experience in developing and testing of such detectors.

**II. Pulsed ionization chamber**
II.-1 Design and experimental set up

Our first attempt was to develop a very simple gaseous detector of alpha particles operating in air in pulse counting mode at gain of one and study its advantages and disadvantages. As was mentioned above, for this particular application operation in pulse counting mode should ensure high sensitivity compared to the conventional ionization chamber operating in current mode. For these studies we have developed and tested two designs of detectors: a single- wire and multiwire detectors. The schematic drawing of the single –wire detector is shown in Fig 1. It is a coaxial chamber with the anode (positively charged) wire diameter of 0.2-2 mm and with the changeable cathode cylinder with diameters ranging from 14 to 120 mm. All outer cylinders have holes along the surfaces covered with 3 μm thick Mylar films to which an alpha source $^{241}$Am could be attached. The inner wire was connected to the custom made charges -sensitive amplifier (see Fig. 2) and the HV is applied to the outer cylinder.

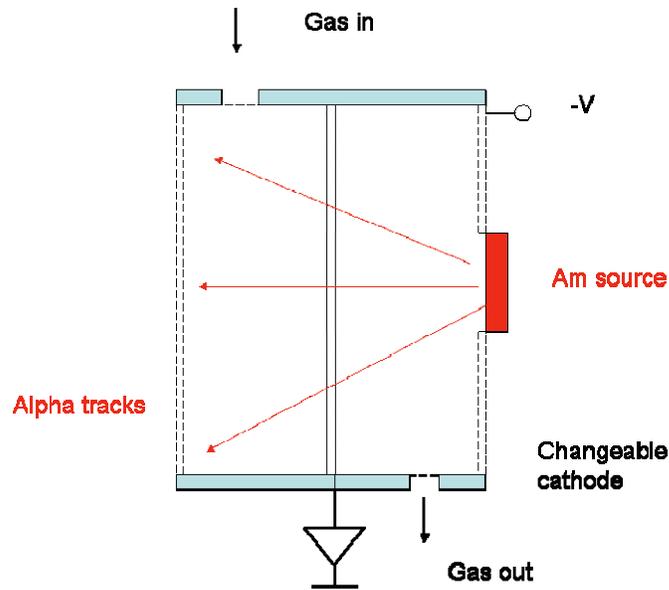

Fig1. A schematic drawing of the single wire pulse ionization chamber

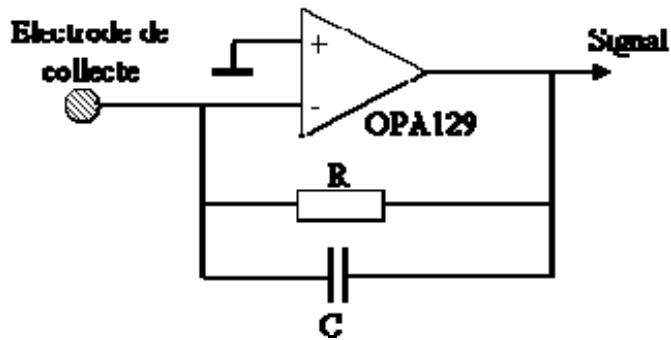

Fig. 2. A schema of the custom made amplifier

The schematic drawing of the multiwire detector is shown schematically in Fig.3. It was a metallic cylinder with diameter of 120 mm inside which in a hexagonal order were stretched inner wires with diameter of 0.2 mm (sensor wires) and 2 mm (H.V. wires). The sensor wires were connected to the charge- sensitive amplifier, whereas the HV wires were connecter to the HV power supply.
 Because our amplifier was designed to deal with slow induced signals, it had a rather high sensitivity to various vibrations including acoustic noise, which affects signal to noise ratio.  To solve the acoustic problem the chamber was suspended inside the metallic shielding box on rubber strings-see Fig. 4.
Both detectors can operate in open air or, if necessary flushed with gas, such as Ar. The gas-flushed detectors were used for evaluation their efficiency with no trapping of the alpha-particle-produced primary electrons by oxygen or other electronegative molecules.

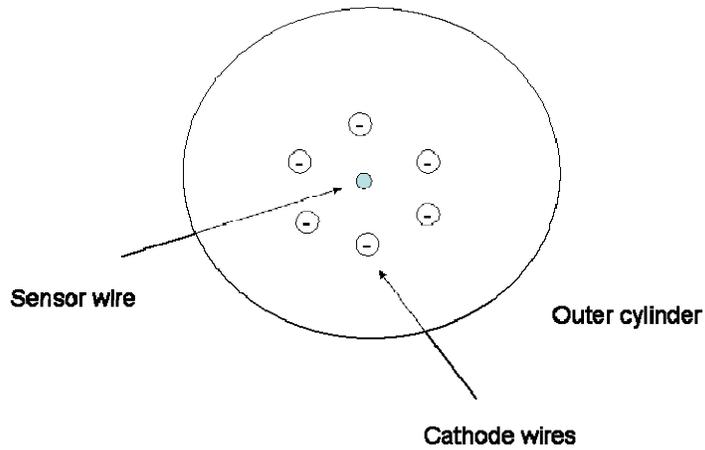

Fig. 3. A schematic drawing of the multiwire pulse ionization chamber. For simplicity only one section of wires arranged in hexagonal order is shown

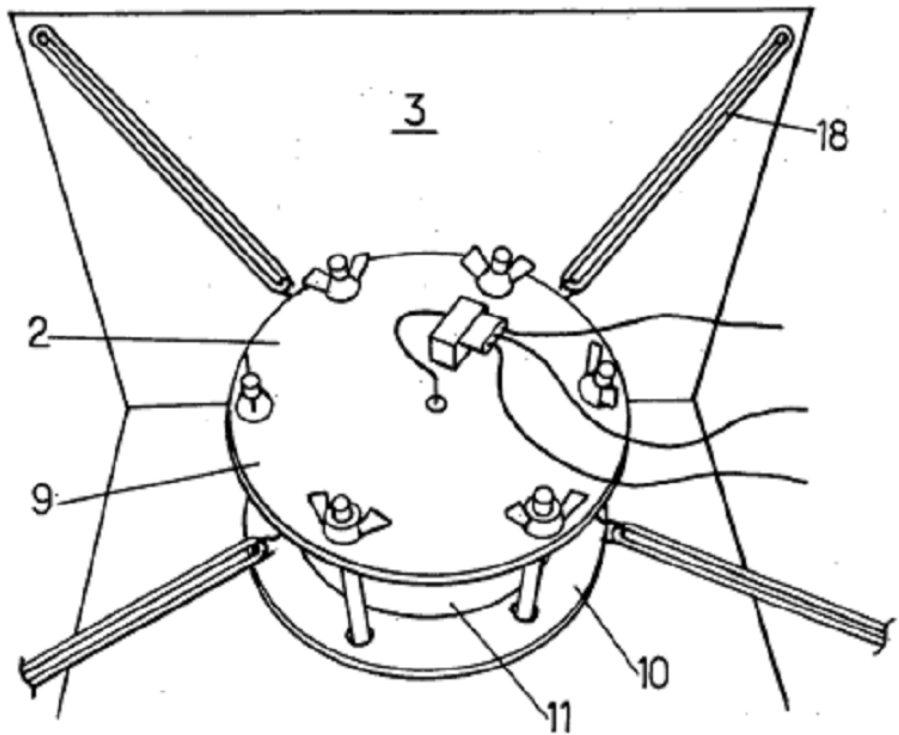

Fig. 4 A schematic drawing of the pulsed ionization chamber suspended on rubber strings

Due to the high drift velocity of primary electrons in Ar the signals produced by alpha particles were easy to detect not only with our custom made preamplifier, but even with a

CAEN commercial charge- sensitive amplifier. After amplification by a research amplifier and discrimination the pulses were counted by a scaler.

II.-2 Test with alpha particles

a) Single- wire detector

The activity of the $^{241}$Am source coated with a 3 μm Mylar film was measured with a custom made dosimeter consisting of a $BaF_2$ scintillator attached to an EMI UV- sensitive photomultiplier . The pulses from the PM were amplified, discriminated and counted by a scaler. One should note that the $BaF_2$ scintillator has a noticeable background counting rate $N_b$ due to the radioactive contaminations inside the crystal. Thus the number of counts/sec produced by the alpha source should be calculated taking into account this background counting rate;
$$N_{alph}=N_m-N_b, \qquad Eq.1$$
where $N_m$ is actual counting rate measured by the PMT when the alpha source was attached to the $BaF_2$ surface.
In such measurements it was very important to correctly adjust the electronic threshold to avoid extra counting caused by 60 keV photons emitted by $^{241}$Am. For this
in the first set of measurements we coated the alphas source with an Al foil, fully stopping the alpha particles, and detected only the $BaF_2$ background pulses and the pulses produced by 60 keV X-rays. This allowed us to set the threshold on the counting electronics to a level when the counting rate due to the 60keV photons was fully suppressed. After this electronic adjustment the Al foil was removed and we measured the counting rate produced by $BaF_2$ radioactive background and by alpha particles $N_m$. These measurements reveal that our alpha source produced $N_{alph}$=132 count/sec.
Since this number was very important for any further calculations we also independently measured the Am source activity by the commercial alpha detector Automess 6150 AD- k. This detector measured ~ 120 count*/s*- close enough to the $BaF_2$ data.
The final set of calibration measurements was performed with a single- wire detector with the cathode diameter of 120 mm flushed by Ar. The alpha source was attached to one of its Mylar windows. Due to the high drift velocity of primary electrons in Ar the short-duration signals produced by alpha particles were easy to detect not only with our custom made preamplifier, but even with the commercial CAEN charge -sensitive amplifier. After amplification by a research amplifier and discrimination the pulses were counted by a scaler. The counting rate produced by alpha particles and by the natural radioactive background was 145 c/s and the background counting rate (the alpha source was covered by an Al foil) was 3c/s, and this allowed us to conclude that the alpha particles counting rate was ~$N_{Ar}$=140c/s.
This value is even slightly higher than one measured with the $BaF_2$ scintillator and with the Automess dosimeter so we assumed that that our detector flushed by Ar has 100% for alpha particles.

After these calibrations we performed measurements with single- wire detectors having various cathode diameters D and all filled with ambient air. Of course, in the case of the air filled detector the "fast" preamplifier becomes insensitive to slow signals on anode wires induced by negative and positive ions, so all further measurements were performed with the "slow" custom made charge-sensitive preamplifier. As an example, Fig. 5 shows the alpha particle signal detected by this amplifier at the condition of an exceptionally very low level of acoustic noise and Fig 6. shows the pulse-height spectra.

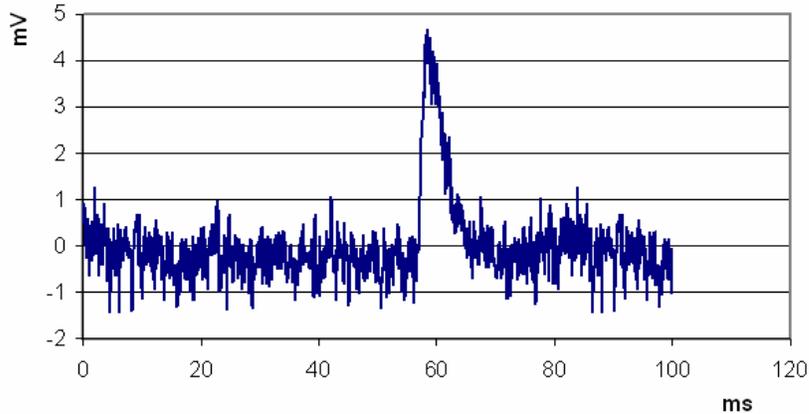

Fig. 5. A pulse from the custom-made preamplifier produce by an alpha particle.

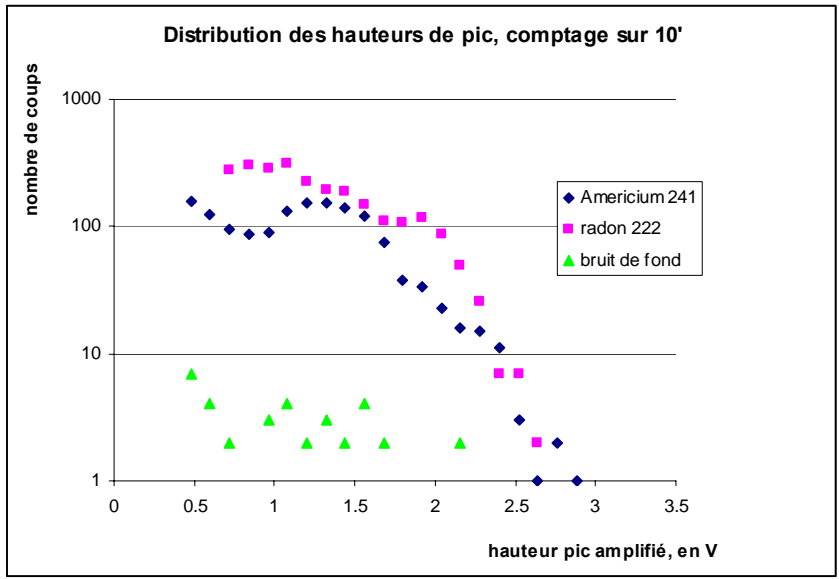

Fig. 6. Pulse-height spectra measures with the sing-wire pulse ionization chamber operating in air

The efficiency of the pulse ionization chamber to alpha particles can be defined by the following relation:

$$\xi = \{N_{air}(D) - N_n(D)\}/N_{Ar}, \qquad eq.2$$

where $N_{air}(D)$- the counting rate measured with the single- wire counter having the cathode diameter of D and $N_n(D)$ is a counting rate when the Am was covered by an Al foil.

The measurements were performed at two different conditions:
a) At a constant voltage V=const= -6kV applied the cathode cylinder
b) At constant electric field near the anode wire E=const (corresponding to -6kV applied voltage at D=20mm; at large diameters the applied voltage was logarithmically increase with D).

The results of the measurements are presented in Fig.7. One can see that the maximum achievable efficiency for alpha particles detection in air is around 13-14.5 % and the efficiency curve is more "flat" in the case when the constant electric field was kept near the anode wire. Thus for detection of alpha particles created in the gas volume (for example in the case of the radon detection in air) it will more favorable to use detectors with diameter up to 120mm and apply rather high voltage:~ 8kV

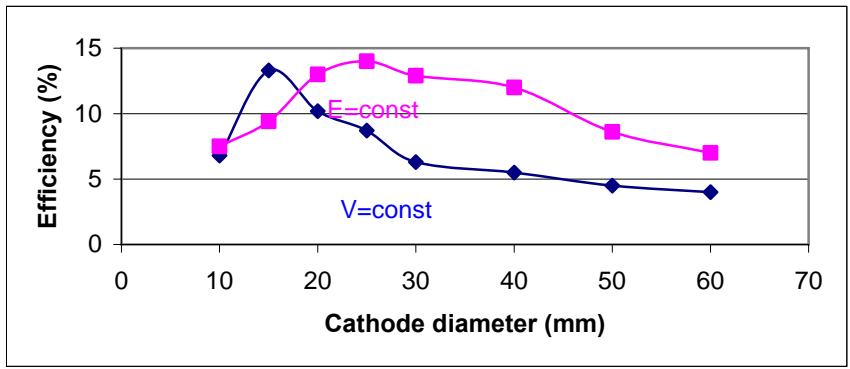

Fig. 7. Efficiency of the single-wire ionization chamber as a function of the cathode diameter D (the anode diameter was 0.25mm) measured with $^{241}$Am source.

One should note that in the case of thin wires (<1 mm) the signal to noise ratio began a sharp increased at V>4kV presumably due to the appearance of gas multiplication.

b) Multiwire detector

Coming from the results presented in Fig 7 it is clear that the single- wire detector reaches the efficiency of ~15% for alpha particle detection only at rather high voltages applied to the changeable cathode cylinders. In the case of the multiwire detector the effective diameter of the cathode (formed by hexagonally arranged cathode wires) was ~30 mm, however the anode wires used had rather small diameters (0.2 mm), so to avoid discharges the efficiency measurements were performed at 4.5kV at the conditions when one or several anode wires were connected to the charge sensitive amplifier.
The results presented in Fig 8. One can see that the efficiency first increased with the number of wires connected, reached the maximum in the case of three wires and then start declining. This drop of the efficiency was obviously due to the increase of the detector capacity; indeed we observed that the signal amplitude start dropping when more than 4 wire were connected to the charge-sensitive amplifier whereas the acoustic noise strongly increased. Probably better results can be obtained if several amplifiers are simultaneously used or even when each wire is connected to a separate amplifier. However, the 22 % efficiency can be sufficient for some applications whereas the cost and complication of the detector may be unnecessary increased with the number of amplifiers used. Certainly the detector design can be further optimized depending on a particular application.

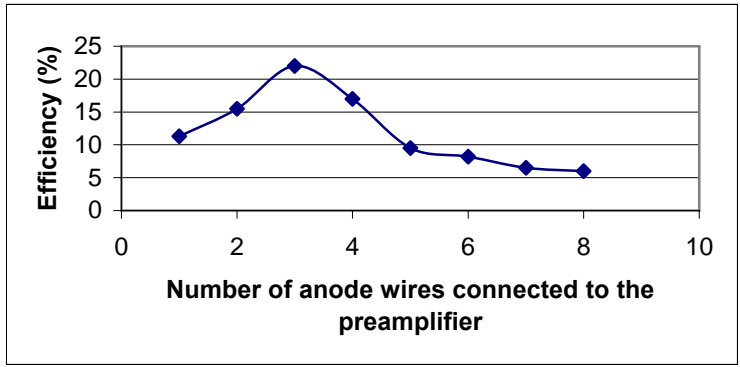

Fig.8. Efficiency of multiwire ionization chamber measured with [241]Am alpha source.

### III. Operation in air in avalanche mode

As is evident from the results presented above, the efficiency of the alpha particle detection by the pulse ionization chamber is below 100% due to the low signal- to -noise ratio.
The detector requires very good isolation measures against vibration and acoustic noise. Such protective measures were successfully developed and implemented (see the [4], but makes the detector rather bulky and fragile, which restrict its applications outside the Laboratory.
To increase the signal to noise ratio we tried to exploit the effect of gas multiplication in air. One should note that the avalanche development in air was quite well studied (see for example [5]). It was discovered that operation in air is not very stable and initial avalanches may easily transit to discharges via a photon feedback mechanism. Probably for this reason the multiplication in air was not exploited so far in any practical device. Indeed, our own tests with single- wire chamber operating in air with gas gain reveal that gains close to $10^3$ can be achieved with gamma radiation ($^{60}$Co was used); however the operation at this gas gain was unstable
Recent developments of hole -type multiplication structures [6-8] open new possibilities; indeed in these detectors due to their geometric feature the photon feedback was strongly suppressed. This is why our further experiments and developments were focused on hole-type multiplication structures.

III-2. Experimental set up and the detector's design

Our experimental set up is shown on Fig. 9. It contains a gas chamber inside which various hole-type multiplication structures can be installed and tested. The distance between the drift mesh and the hole-type multiplier can be varied from 1 to 4cm. The gas chamber can be flushed with various gas mixtures or filled with ambient air.

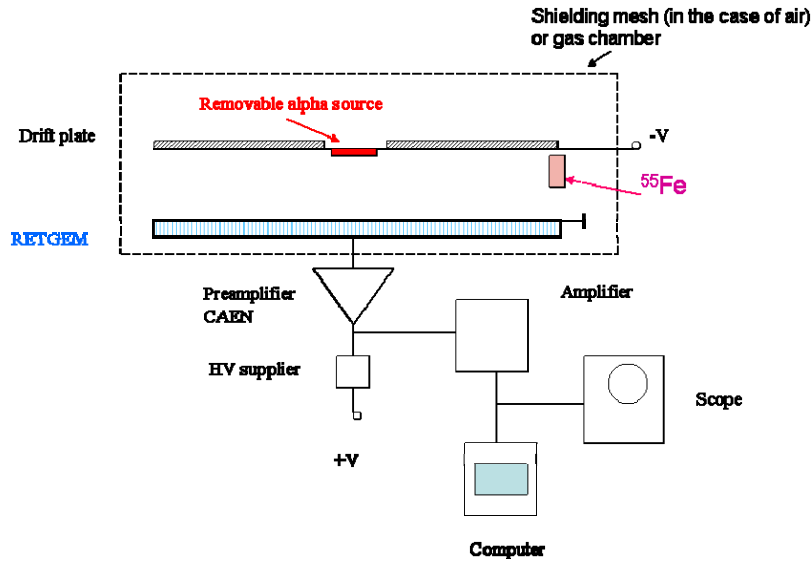

Fig. 9. A schematic view of the experimental set up used for studies of the RETGEM operation in air. Typically the upper electrode of the RETGEM was grounded and positive voltage was applied to its bottom electrode

The ionization inside the chamber was produced by an $^{241}$Am source (one should note that the Am source used in these measurements was not the same as was used the in the case of the pulsed ionization chamber) or by $^{55}$Fe. Each of the sources can be closed by special shutters, which allowed us to verify that observed signals were caused by these sources, but not by spurious pulses. Because the signals from the hole-type multiplication structures were rather fast[1] actually in any gases used including air, a conventional Ortec or CAEN charge- sensitive amplifier were used for their detection. The pulses after the amplification by a research amplifier were then sent to a personal computer where they were treated and viewed by a LabView program.

Our first attempts to achieve multiplication in air were done with a GEM [8] and also with so called "optimized" or "thick" GEM (TGEMs) [9], however both these detectors, especially the GEM were not able to operate stably for a long time in ordinary, ambient air containing dust particles. To solve this problem we invested quite a lot of efforts to develop more robust version of TGEM having Cr electrodes coated with thin CrO layers- see Fig 10 [10]. The important feature of this detector was dielectric rims 0.1-0.2 mm thick manufactured by photolithographic technology around holes. We called this detector Resistive Electrode TGEM or RETGEM. Dielectric coating together with dielectric rims made this detector extremely robust enabling it to operate in air containing dust[2]. RETGEMs with the following geometrical characteristics were used in this work: thickness 0.5 -1mm, diameter of holes 0.3-0.5mm , pitch 0.6-0.8mm.

---

[1] The reason that the pulses were fast in this case is that the electric field inside the holes is rather uniform and quite high in contrast to wire detectors in which the electric field decrease inversely as distance from the wire.

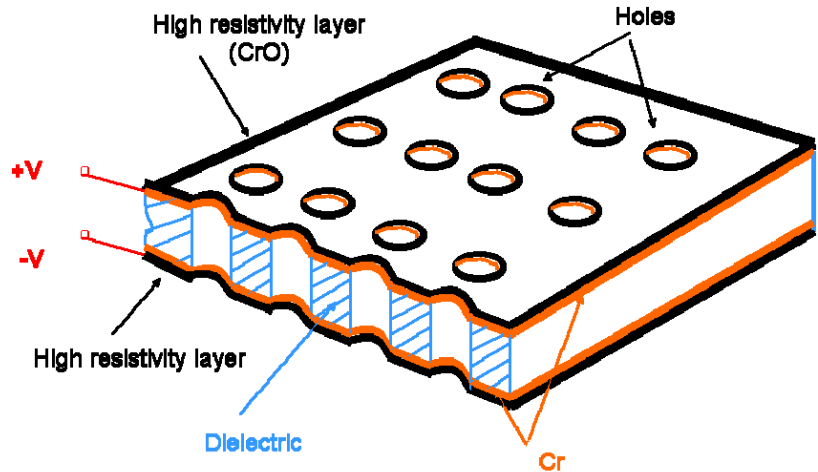

Fig. 10. A schematic drawing of a RETGEM with double layer Cr/CrO electrodes

The experiments and tests described below demonstrated that an air filled detector based on RETGEM (single RETGEM or two RETGEMs operating in cascade- see below ) was a very successful device allowing to detect not only alpha particles in air with 100% efficiency, but also soft x-rays, for example 60 keVx-ray from $^{241}$Am. Moreover, RETGEM can be used in combination with other detectors, for example MICROMEGAS and this double multiplication structure also can operate stably in ambient air.

III-2. Results obtained with RETGEMs
  a) Single RETGEM

Because of the oxide- coated RETGEM is a novel device, we made careful studies of its operation. First tests were done in gases and gas mixtures in which ordinary GEM and RETGEM were earlier studied. These comparative studies allowed us to better understand the properties of the RETGEM. The final tests of course were done in ambient air. Some of our results are presented on Figs 11 and 12, showing the signal amplitude (raw data) from the research amplifier vs. the voltage applied to the RETGEM electrodes in various gases: Ar, Ar+10%$CO_2$. Fig. 11corresponds to the RETGEM 1mm thick with holes diameter of 0.3 mm whereas Fig 12 shows the data obtained with the RETGEM having 0.5mm holes. Our charge -sensitive amplifier was calibrated by the charge injection method so from the amplitude of signals in Ar and Ar+ $CO_2$ one can calculate the gas gain. For example in the case of the $^{55}$Fe a 10V signal corresponded to a gain of ~$10^4$. One can see that signal amplitudes of 0.5 -1V were measured with alpha particles in air. However, the gas gain in air could not be calculated so easily. It is known that in the case of electronegative gases [11], including air, only negative ions reach the hole- type structure ( all primary electrons created by alpha particles are captured by oxygen molecules). However, in the strong electric field inside the holes some of these electronegative ions lose the attached electrons (by so called an electron detachment effect) and these free electrons Townsend avalanches. A priory it is not known precisely how many electrons were detached so one cannot calculate the gas gain simply from the measured signal amplitude as was possible in the case of Ar and Ar+$CO_2$. For the gain evaluation we used a special procedure described in the next paragraph. Actually, what was important in these first preliminary tests was to determine the detector efficiency for alpha particles in air. As in the case of the pulse ionization chamber described above, we determined the efficiency as a fraction of alpha counting rate, measured with RETGEM in air and Ar;

---

[2] Although dust particles on the dielectric rims due to their polarization cause filed line concentration on their sharp surfaces this does not provoke discharges since there are no any conductive pass for current to flow

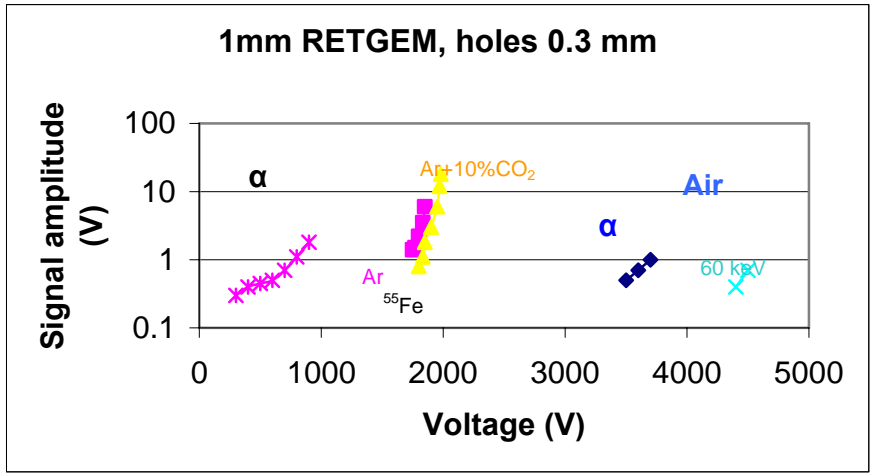

Fig. 11. Signal amplitude vs voltage measured with RETGEM (holes 0.3mm) in Ar (alpha particles and 6keV photons), Ar+10%$CO_2$ (6 keV photons) and in air (alpha particles and 60 keV photons)

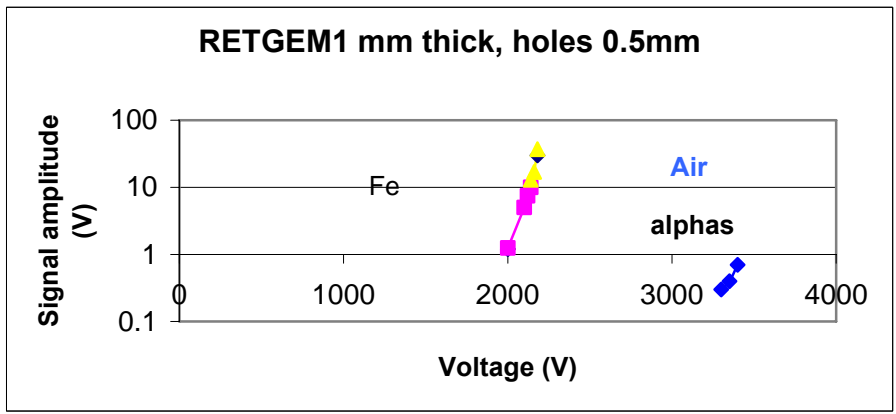

Fig. 12. Signal amplitude vs. voltage measured with RETGEM (holes 0.5mm) in Ar (rose) and in Ar+10%$CO_2$ (yellow) with 6 keV photons and in air (blue) with alpha particles

$$\xi_{RETGEM}=\{N_{air}-N_{nair}\}/N_{Ar} \qquad \text{eq.3},$$

where $N_n$ air is the counting rate in air when the alpha source was closed by Al foil. Figs 13 and 14 show some results of such measurements in Ar and air respectively.

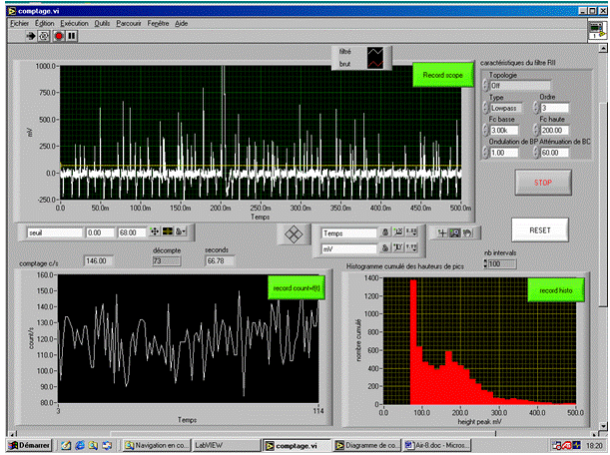

Fig 13. A print of the LabView screen showing some results of measurements with alpha particles in Ar at pressure of 1 atm: the analog signals from the RETGEM (top screen), counting rate vs. time (left bottom screen) and a pulse-height spectra (right bottom screen)

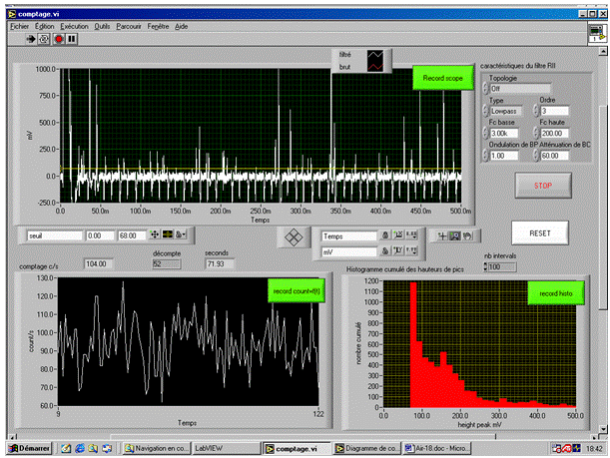

Fig. 14. Results obtained with a single stage RETGEM operating in air. The voltage across the RETGEM was 3200V, the thickness of the drift region 2cm

These figures show a copy of LabView screens. The upper part of the screens show the analog signals from the RETGEM. On the bottom on the Lab View screen are shown: on the left- the counting rate measurements and on the right -the pulse height spectra. One can see that the alpha counting rate in Ar was $N_{Ar}$=120Hz, whereas the counting rate from the RETGEM operating in air was $N_{air}$=95, thus the detection efficiency was 80%- almost 4 time higher that in the case of pulse ionization chamber ($N_{nair}$ and $N_{nAr}$ were~0).

Note that the RETGRM-based detector was not very sensitive to vibrations and thus can be used in more harsh conditions than the pulsed ionization chamber.
During the long term tests of RETGEM operation in air it was observed that in days when the ambient air humidity was high (>30%) spurious pulses may appear. These pulses disappeared when the gas chamber was flushed with dry air. To suppress this undesirable effect we modify our device by adding a second RETGEM operating in tandem with the first RETGEM.

b) Double -stage RETGEMs

The schematic drawing of the double RETGEM is shown on Fig 15. The HV to the detector's electrodes were applied using a resistive divider. Systematic tests of this detector show that in contrast to the single RETGEM, the double RETGEM could operate in ambient air much more stably-even at the humidity level close to 70%. Presumably this was because the voltage drop of each RETGEM in two-stage configuration was less than in the case of the single RETGEM and thus the charge leaks due to the humidity was less. In addition, if necessary, higher amplitude of the output signal can be achieved before the sparking appeared. Thus we consider double RETGEM as being more suitable for a practical alpha detector operating in air.
For curiosity we also flushed our detect with a 100% humidity air. In this case a leak current appeared between the detector electrodes, however in 10 min after the flushing was stopped the detector resumed a normal operation.
Figs. 16 and 17 show the LabView screens corresponding to the measurements with the double RETGEM in Ar and air restively. From comparing the measured counting rates one can see that the efficiency of ~100 % was achieve in air with the cascaded RETGEM.

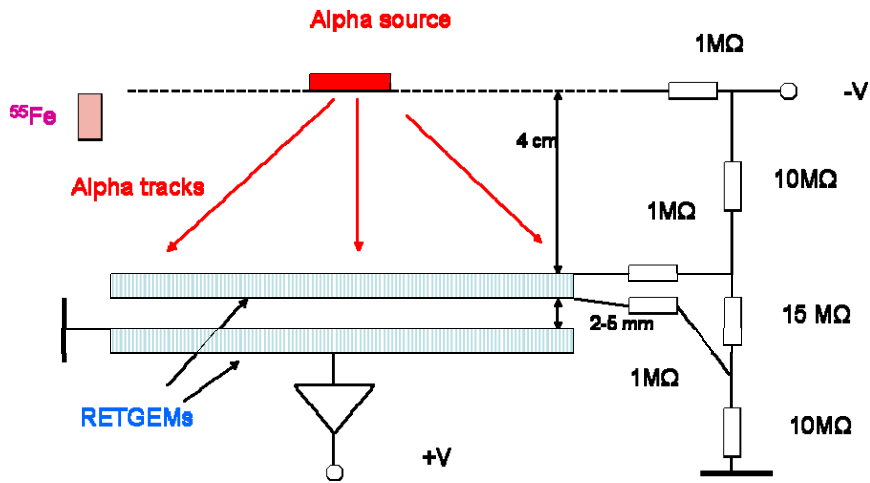

Fig. 15. A schematic view of the double RETGEM feed with voltages via a resistive divider chain

The results of the detection efficiency measurements vs. the applied voltage is shown in Fig. 18. One can see that a very good counting plateau was achieved.
The other important measurements were done to determine if this higher efficiency was obtained because some alpha tracks in the set up we used directly hit the top RETGEM. To verify this we changed the position of the source as shown on Fig. 19: the alpha source was placed 2 cm above the top RETGEM and its active part faced the drift mesh. In this geometry alpha tracks could not hit the RETGEM surface. The detector efficiency measured in this condition also was 90% indicating that electronegative ions could drift a distance more than 2 cm before entering the holes of RETGEMs where the electron detachments and avalanche development further occur.

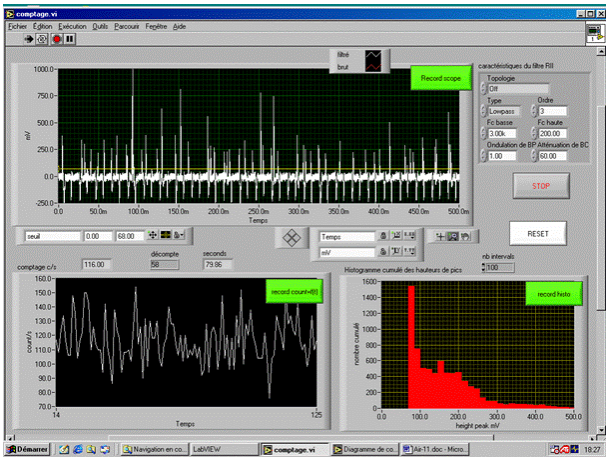

Fig. 16. Results obtained with a two- stage RETGEM detecting alpha particles in Ar. Detection efficiency for alpha particles was 100%

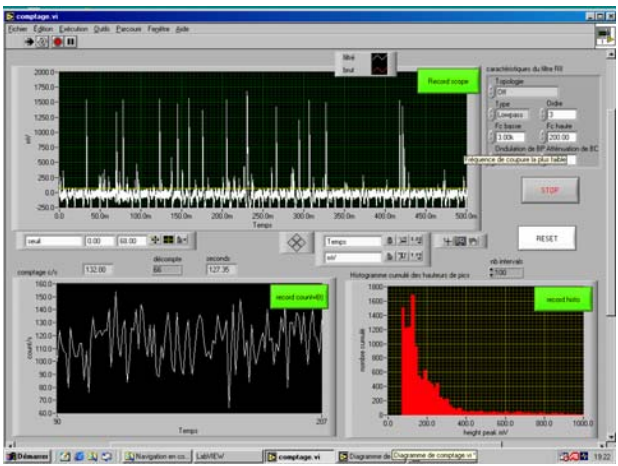

Fig. 17.Results obtained with double RETGEM detecting alpha particles in air. The voltage on the resistive divider was -6kV, the voltage across the bottom RETGEM was 3.6kV.From the counting rate value one can see that efficiency of alpha particles detection was 100%.

Finally we performed measurements allowing rough gas gain estimation for the double RETGEM operating in air. For this the gas chamber again was filled with Ar and we operated the detector at gain $A_{Ar}=10$. The signal amplitude $S_{Ar}$ at this condition is equal:

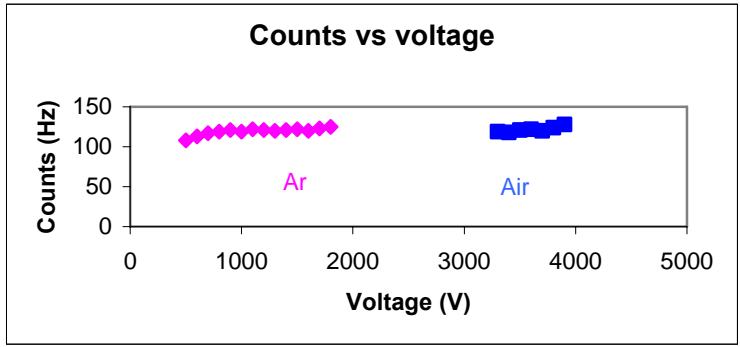

Fig. 18. Counting rate vs. voltage applied across the bottom RETGEM measured in Ar and air

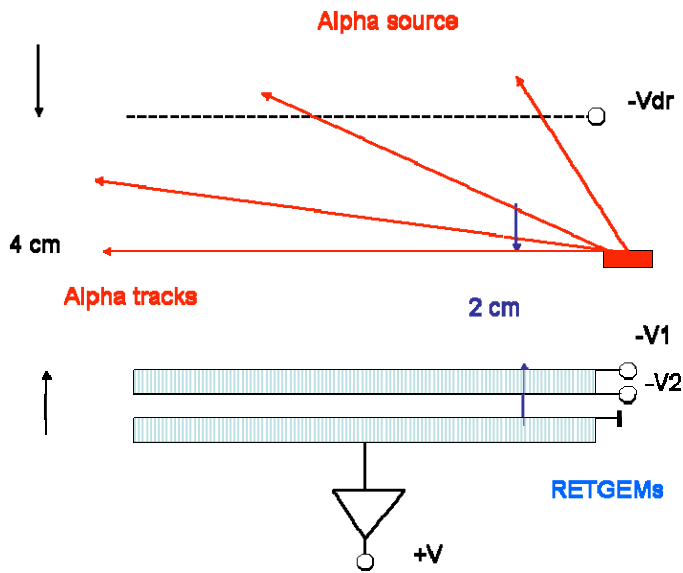

Fig.19. A schematic drawing of the measurements with the alpha source facing the drift mesh Efficiency of 90% was achieve in this geometry. Thus electronegative ions can be drifted at least 2 cm in air and then trigger avalanched in the RETGEM holes

$$S_{Ar}=kA_{Ar}n_0 \qquad \text{eq.4,}$$

where k is a coefficient and $n_0$ is the number of primary electrons produced by alpha particles.
The number of primary electrons can be calculated
$$n_0=B/W_i \qquad \text{eq.5 ,}$$
where B is energy of alpha particles released in the gas and $W_i$ is a mean energy necessary for Ar ionization.

The signal amplitude measurements vs. time corresponding to this case (the first 10 min) are shown in Fig.20 ( $S_{Ar}$=4.5V). Then a portion of air was introduced into the chamber. The signal amplitude in Ar+air mixture dropped (the time interval 10-20 min in Fig 20). By increasing the voltage on the voltage divider (see Fig. 15) and on the anode of the bottom RETGEM the signal amplitude was restored to a level of S=3V.
In this case the signal amplitude is equal:
$$S_{mix}=kA_{mix}n_{mix} \quad (5),$$
where $n_{mix}$ is the number of primary electrons triggering avalanches in the holes of the RETGEM. Because a great fraction of primary electrons in the mixture with air is captured by oxygen molecules $n_{mix}<n_0$, thus the gain is $A_{mix}>A_{Ar}S_{mix}/S_{Ar}$ or> 5. Such procedure was repeated a few times till the chamber was fully filled with air. From the rough estimations similar to described above one can expect that the gain achieved in air can be as high as $\geq 10^4$.
From the fact that the efficiency of our detector was 100% one can independently support the statement that gains of $10^3$-$10^4$ can be achieved in air. For example, at mean signal amplitude of S~0.5V ( the starting point for the gas gain curve in air in Fig.12) the collected charge on the detector electrodes was ~$10^5$ electrons, thus that if the gain at this amplitude is ~$10^3$ than ~100 primary electrons triggered avalanches in the detector's holes. However, if one assume that the gas gain at this amplitude is more than $10^4$ then the number of primary electrons is less than 10 and this will immediately (due to the Poisson statistic), affect the efficiency -it will drop, which was not observed. Of course, with the voltage increase the amplitude of the signal and the gain increases too and finally can reach the value of $10^4$. Qualitaively the same conclusion can be drown from the poor energy resolution obtained in air.
To independently verify that gains $\geq 10^4$ can be achieved, we tried also to detect 60kev X –rays produced by a $^{241}$Am source. For this the Am source was covered by Al foil fully stopping the alpha particles and indeed at voltages >4200V X-ray pulses were clearly observed - see curves marked "X-rays" in Fig 12.

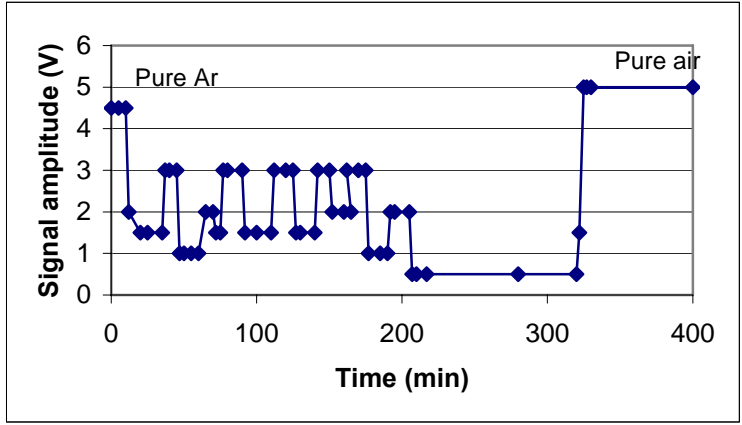

Fig. 20. Alpha signal variation with time when air was gradually introduced to the gas chamber initially filled with pure Ar.

c) RETGEM+MICROMEGAS

As was mentioned above, the main reason why the hole- type detector can operate stably in air is because in this geometry the photon feedback from the cathode is strongly suppressed. On the other hands, our earlier studies showed that that the ratio of the number of emitted photos from the Townsend avalanche to the gas gain drops with electric filed [12], suggesting that in detectors having very high electric filed in the amplification gap the light emission (for the given gas gain) will be less.
An example of such detector could be MICROMEGAS which has a very strong electric field in the multiplication gap. So it was interesting to check if with MICROMEGAS one can achieve higher multiplication in air compared to single -wire counters or ordinary parallel-plate chambers. To answer this question we installed MICROMEGAS in our gas chamber (see Fig. 21) and performed measurements with alpha particles in various gases including ambient air. Some results are shown in Fig. 22. One can see that signal amplitude in the interval 0.5-1 V were observed

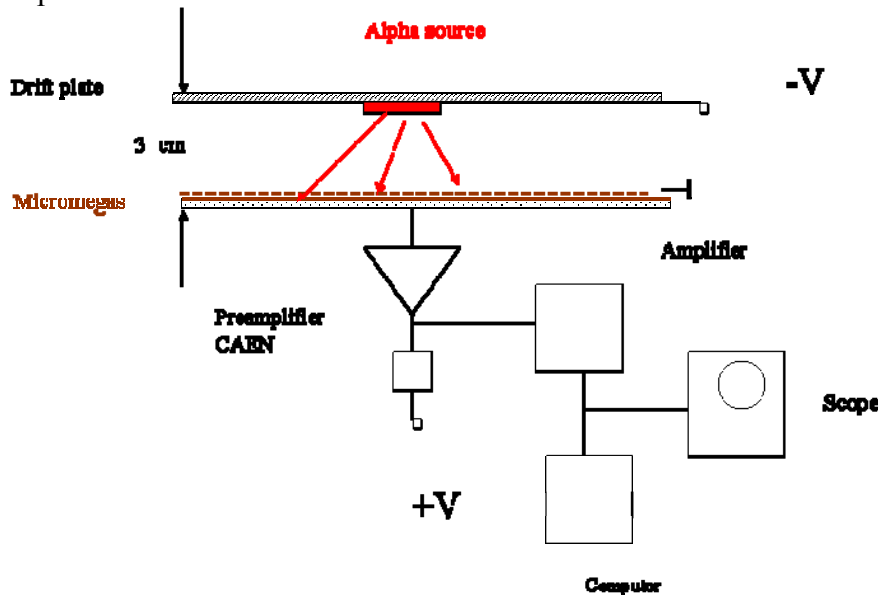

Fig. 21. A schematic drawing of the experimental set up for test of MICROMEGAS operation in air

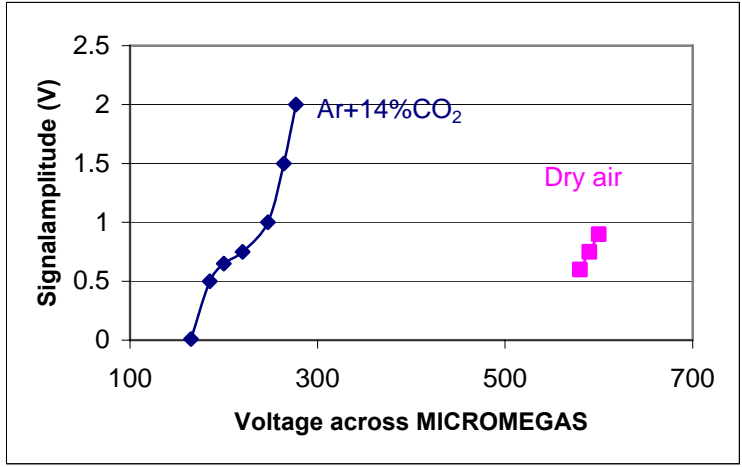

Fig. 22. Signal amplitude vs. voltage applied across MICROMEGAS measured in Ar+14%$CO_2$ and in dry air

in air with MICROMEGAS. Unfortunately the MICROMEGAS operation
in ambient air was unstable; large amplitude spurious pulses often appeared triggering discharges in the MICROMEGAS. We attributed this unstable operation to the charge leaks across the MUCROMRGAS spacers caused by humidity. Indeed in dry air the MICROMEGAS operation was much more stable.
To solve this problem of unstable operation we assembled and tested MICROMEGAS operating in cascade mode with the RETGEM placed a few mm above the MICROMEGS and used as a gas gain booster. Indeed, a stable operation was achieved with this detector configuration. In Fig. 23 are show gain curves for single RETGEM and RETGEM combined with MICROMEGAS, from which one can conclude that MICROMEGAS allowed to boost the overall gain by a factor of 5. An important advantage of this approach is that the overall voltage applied to such two-stage detector is lower that in the case of the double RETGEMs which can be attractive in practical device.

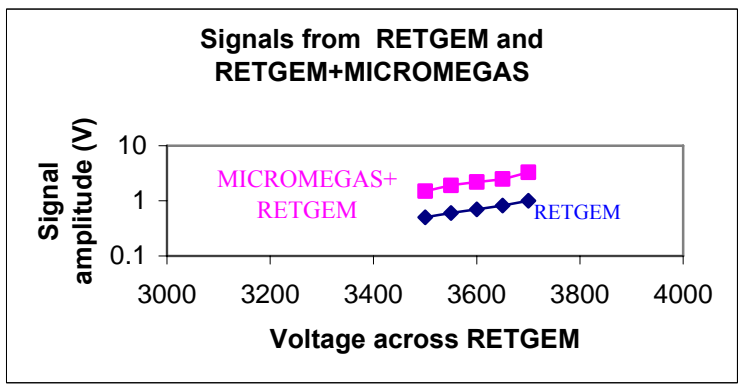

Fig. 23. Amplitude of the alpha signal vs. the voltage applied across the RETGEM (blue) and in the case of RETGEM operating in cascade with MICROMEGAS (rose)

**Discussion**

Two types of alpha particle detector operating in ambient air in pulse counting mode and /or with gas multiplication were developed and tested in this work. The first one is a pulsed ionization chamber able to detect with 15-22% efficiency individual pulses produced by Am alpha particles. This relatively low efficiency was due to the sensitivity of our custom made amplifier to various vibrations and acoustic noise. The single -wire version of this detector was already used for detection of Rn in ambient air. As an example Fig. 24 shows results of the Rn measurements in our laboratory at Ecole des Mines in St. Etienne and in the basement situated just under the Laboratory. One can clearly see the increase of the Rn concentration on a factor of five in the basement. One should note that in contrast to Am, which was attached at a well defined place on the detector cathode cylinder, the Rn alpha particles are randomly distributed over the detector volume and in many occasions their signals were below the electronic threshold of the counting system being selected high enough to cut parasitic acoustic noises. As a results the sensitivity of our single -wire detector to Rn was 40Bq/m3 for 10 min measurement time and 7bq/m3 for 6 hours measurement time which is a factor three lower than some commercial device have (see for example specification for ATMOS [13]). The pulse- height spectra of the alpha particles produced by Rn is shown in Fig. 6; as one can see the energy resolution achieved with a single –wire detector is much worse compared to some commercial devices (see for example [13]). However, we think that our device can be manufactured cheaply and thus it still can compete in some applications with high performance, but more expensive commercial devices. Moreover, the multiwire version of this detector has potential for better efficiency and better energy resolution.

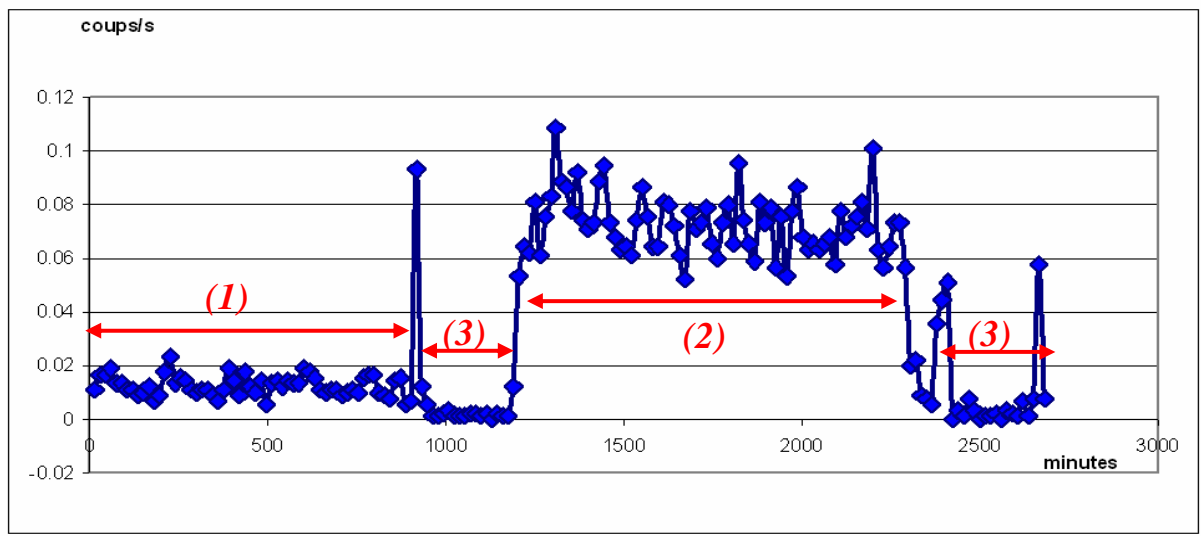

Fig. 24. Results of measurements of the Rn background with a single-wire pulsed ionization chamber: 1-Counting rate measured in the Laboratory 2-counting rate measured in the basement, 3-the single wire counter was flushed with fresh air taken outside the building.

However, there is another way to improve the signal to noise ratio- to exploit the gas multiplication mode. This possibility was successfully demonstrated with the second type of the air filled detector - RETGEM. This detector allowed easily achieved ~100% efficiency with Am alpha particles.

We consider that this type of the detector should be useful not only for Rn detection, but also for detection of alpha particles emitted by surfaces (surfaces alpha contamination). Indeed as it was shown above, with the RETGEM we were able to detect even 60 keV energy deposited in the air volume which is equivalent to detecting only a small fraction of the alpha track. Because of the mean free path of alpha particles in air is around 4 cm, one can scan surfaces with this device moving on a distance of a few cm from them-see Fig. 25. Moreover it was demonstrated the negative ions can be drifted in air at leas a few cm and this effect can be exploited in practice in order to build a long -range alpha particle detector. In connection to this one should note that the authors of recent work [11] succeeded to build a TPC prototype for high energy physics experiments in which electronegative ions were drifted 8 cm before they entered the readout gas amplification structure .This results indicate that in clean enough ambient air (not much dust or aerosols) one also can drift electronegative ions on distance more that a few cm, which can make a practical device more convenient in exploitation.

Of course, the energy resolution of our present prototype is much below the best commercial devices, but due to its estimated low cost it can be massively used in some areas which may require continuous monitoring of alpha particle contamination (for example Po): airports, railway stations and so on. In the present version the RETGEM can be used as a trigger of the "first level" alarm in these areas assuming that more refined analysis can be done a few minutes latter with a more powerful and expensive alpha analyzer.

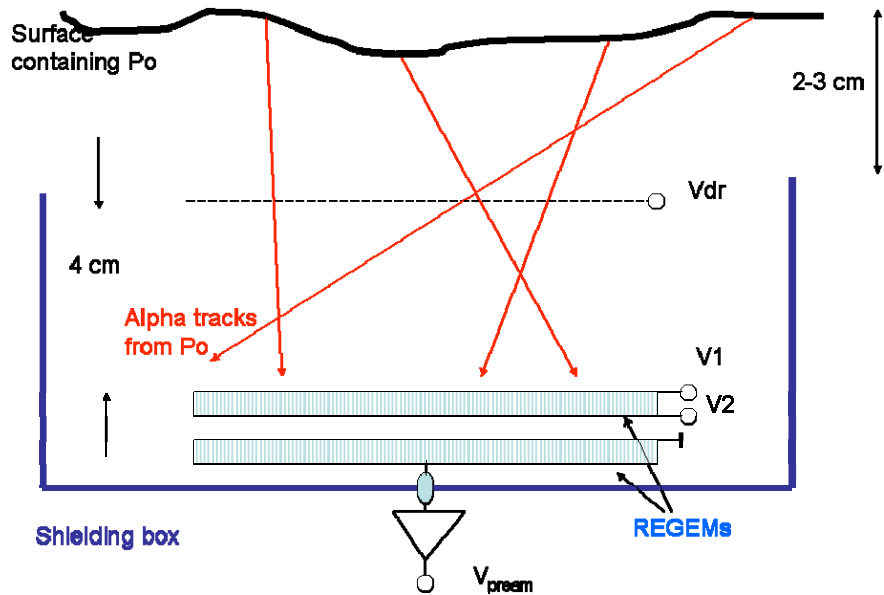

Fig.25. A possible design of the detector for Po monitoring. As follows from our studies for the "short-range" operation mode the voltage applied to the detector electrodes should be: Vdr~-6kV, $V_1$~-4.3kV, $V_2$~-1.7 kV and the voltage applied via the charge-sensitive amplifier " $V_{pream}$"~+3,6kV. In the case of the "long-range" detection mode all applied voltages should be positive in order to ensure the drift and collection of electronegative ions created in the gap between the investigated surface (grounded) and the drift mesh. Thus the expected voltages are: Vdr=+1kV, $V_1$~+2.7 kV, $V_2$ ~5kV , the top electrode of the bottom RETGEM should be disconnected from the ground and kept at +6kV and finally ~9.6 KV should be applied to the bottom electrode of the bottom RETGEM.

## Acknowledgments.

Authors would like to thank Prof. Richard Garvin for very useful discussions, suggestions and advices through out the work and J.-C. Santiard, who design the charge-sensitive amplifier and participated in some measurements.